\documentclass[12pt]{iopart}

\usepackage{amssymb}
\usepackage{amsbsy}
\usepackage{graphicx}
\usepackage{color}

\newcommand{\be}{\begin{equation}} \newcommand{\ee}{\end{equation}}
\newcommand{\bea}{\begin{eqnarray}} \newcommand{\eea}{\end{eqnarray}}

\begin{document}
\begin{flushleft}
       \hfill                       UAB--FT--569\\ \hfill
      July 2004\\ \hfill \\
\end{flushleft}
\title{Bose-Einstein Condensation, Dark Matter and Acoustic
Peaks}

\author{F. Ferrer\dag~and J.A. Grifols\ddag}

\address{\dag Astrophysics, University of Oxford\\
Denys Wilkinson Building, Keble Road, Oxford, OX1 3RH, UK}

\address{\ddag Grup de F\'\i sica Te\`orica and Institut de F\'\i sica
     d'Altes Energies\\ Universitat Aut\`onoma de Barcelona, 08193 Bellaterra, Barcelona, Spain}

\eads{\mailto{f.ferrer1@physics.ox.ac.uk}, \mailto{grifols@ifae.es}}

\begin{abstract}

\noindent
Scalar mediated interactions among baryons extend well above the Compton wavelength, when they are embedded in a Bose-Einstein condensate composed of the mediating particles. Indeed, this non-trivial environment results in an infinite-ranged interaction. We show that if the Dark Matter of the Universe is composed of such a condensate, the imprints of an interaction between baryonic and Dark Matter could be manifest as anomalies in the peak structure of the Cosmic Microwave Background. 

\end{abstract}


\tableofcontents

\setcounter{secnumdepth}{3}

\section{Introduction}

 In work carried out over past years, we have studied dispersion
forces~\cite{Feinberg:ps} (i.e. van der Waals type forces) exerted
among macroscopic bulk matter mediated by light(or massless) particle
exchange, both in vacuo~\cite{Grifols:zz,Ferrer:1998ue} and in a heat
bath~\cite{Ferrer:1998ju}.  Known examples are the Casimir-Polder
forces~\cite{casimir} among neutral atoms (where the exchanged
particles are photons) or the Feinberg-Sucher two-neutrino exchange
forces~\cite{sucher} among baryonic matter. Because real matter sits
in the cosmic microwave background and in the relict neutrino
background, those forces are affected by the photon thermal background
and the neutrino background,
respectively~\cite{Ferrer:1998ju,Barton:2001ej}. Now, scalar particles
are fundamental ingredients of the Standard Model of Particle Physics
(SM) and completions thereof. Known examples are, Higgs bosons,
axions, majorons, scalars appearing in supersymmetric extensions of
the SM, dilatons, radions, etc. Hence, when coupled to ordinary
matter, their exchange leads also to dispersion-like forces~\cite{Ferrer:1998rw}. 

Scalar particles, such as the axion, have been considered in the past
as candidates for the dark matter in the Universe. While the favorite
candidates for the dark matter in the Universe are weakly interacting
particles with masses in the $GeV$ range (WIMPS), the most popular
example of this being the lightest supersymmetric particle (in
particular, the neutralino)~\cite{Ellis:1983ew}, it cannot of course
be excluded that a scalar (even one not given in the list above) is
actually the real ingredient of dark matter~\cite{Matos:2000ki}. In this respect, it has
been recently proposed that the observed $511\; keV$ emission from the
Galactic bulge~\cite{Jean:2003ci} could be the product of very light
annihilating scalar dark matter particles~\cite{Boehm:2003bt}.

Therefore, the putative
forces caused by the double exchange of such scalars would be also
affected by the presence of the scalar dark matter. When two scalar
particles are exchanged in the t-channel, spin-independent dispersion
forces are being generated that add coherently over unpolarized bulk
matter and extend over distances on the order of the Compton
wavelength of the mediating scalars. If the fundamental coupling of
these scalars to matter fermions is of the usual Yukawa type, then the
long distance behavior of the associated potential in vacuum is
specifically of the form $\sim
exp(-2mr)/r^{5/2}$~\cite{Ferrer:2000hm}. When the bodies subject to
those forces are embedded in a heat bath made of the same scalars, a
very dramatic effect takes place provided the scalar particles carry a
conserved charge and the boson gas reservoir is characterized by a
nonzero chemical potential. Namely, the finite range potential (now
behaving as $\sim exp(-2mr)/r^2$) becomes infinitely ranged when the
heat reservoir suffers Bose-Einstein condensation. Specifically, an
$\sim exp(-2mr)/r^2$ potential turns into $\sim 1/r$. Conversely, when
the heat bath makes the transition to the uncondensed phase, this
infinite range force becomes finite ranged. The phenomenon, first
described in ref~\cite{Ferrer:2000hm}~(see also~\cite{Consoli:2003jv}), comes about as a combination of
kinematics (three-momentum exchange of the matter system with the
medium) and the collective effect of condensation of charge. In this
paper we wish to explore a physical realization of this effect. Our
purpose is to display a system where the phenomenon is of physical
import.  Since our primary goal is to investigate the implementation
and physical consequences of Bose-Einstein condensation of a
relativistic scalar gas itself, we will not focus on any specific
particle physics model (see e.g.~\cite{Sannino:2003mt} for the effects on the Standard Model particle dispersion relations) nor claim that the envisaged cosmological arena
is a realistic one either. We consider it as a convenient playground
for testing the potentialities of a model characterized by matter in a
light scalar heat bath. 

On the other hand, one should be open minded as to which is the actual
nature of dark matter and consider other options as long as WIMPS are
not experimentally detected~(see e.g.~\cite{Bertone:2004pz} for recent accounts of the status of particle dark matter). Following the progress of experiments in measuring the cosmological parameters, the so-called $\Lambda$ CDM scenario has emerged as the {\it Standard} cosmological scenario~\cite{Krauss:2004iq} suggesting that most of the energy density of the Universe is stored in a dark sector, $2/3$ of which, dubbed dark energy, has negative pressure and the rest is composed by pressureless dark matter~(although see~\cite{Blanchard:2003du} for alternative fits to the experimental data). The simplest alternatives are a cosmological constant, $\Lambda$, to explain the recent period of accelerated expansion and WIMPS for the dark matter, but other candidates have been put forward that usually entail a richer structure in the dark sector. To name just a few, dynamically evolving scalar fields are at the basis of quintessence models for the dark energy~\cite{Peebles:1987ek}, condensates may play a significant role in many cosmological phenomena~\cite{Enqvist:2003gh} and provide a unified description of dark energy and dark matter~\cite{Bassett:2002fe}, also long range forces in the dark matter sector motivated by string theory have been examined in~\cite{Gubser:2004uh}. 

In section~\ref{sec:model} we introduce the Bose-Einstein condensate of scalar dark matter particles and analytically study their influence on the peak structure of the CMB. A numerical example is next discussed in section~\ref{sec:numeric} before ending with our conclusions in section~\ref{sec:conclusio}.

\section{Bose-Einstein condensate of Dark Matter} \label{sec:model}

\subsection{The set-up}
Let us start by displaying the model. We assume the existence of light
scalars that carry a conserved quantum number Q. These scalars
constitute the bulk of dark matter. For a relativistic boson
gas~\cite{Haber:fg,Kapusta:aa,Bernstein:kf}, i.e. $m/T <<1$, condensation occurs
below $T_c=(3q /m)^{1\over 2}$ for a fixed charge density
$q$. Conversely, for given $T$, when the charge density exceeds
$q_c=mT^{2}/3$, Bose-Einstein condensation follows. Now, in an
expanding Universe number densities vary with $T^3$ and hence there
should be a period in the history of the early Universe where a
condensed phase of scalars coexists with a gaseous scalar phase: for a
sufficiently high temperature, the charge density exceeds
$q_c$~\cite{Bernstein:kf}.  The scalars, while they are in a condensed
phase, are a pressureless fluid and thus constitute cold dark
matter. In order that structure formation can proceed as usual, we
certainly want the scalar background to be in the condensed phase at
matter-radiation equality and beyond, and the eventual transition to
an uncondensed phase to take place not before the large scale
structures we observe today have had enough time to form. We shall see
below how these conditions can be met.

\begin{figure}[htb]
\begin{center}
\includegraphics[clip=true,width=\textwidth]{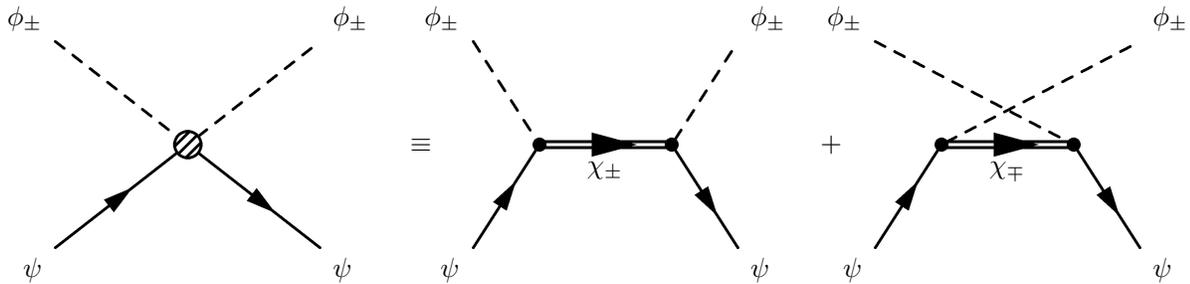}
\end{center}

\caption{Effective coupling of the scalars $\phi_{\pm}$ to ordinary baryons $\psi$.} 
\label{fig:feyn}
\end{figure}

\begin{figure}[htb]
\begin{center}
\includegraphics[clip=true,height=5cm]{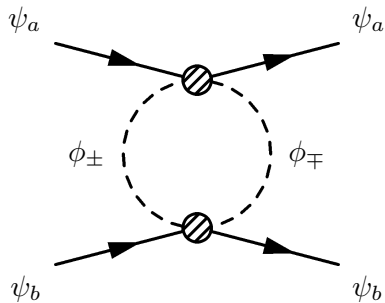}
\end{center}

\caption{Exchange of two scalars giving rise to the force among baryons.} 
\label{fig:feyn2}
\end{figure}

We suppose next that our scalars couple to ordinary baryons via
the effective interaction
\begin {equation}
\label{eq:lagrangian}
{ g_{eff} \over m_{b} }\bar \psi \psi \phi \phi ^{\dag}
\end {equation}
where $m_{b}$ is the nucleon mass, arbitrarily chosen to make the
effective coupling $g_{eff}$ dimensionless. It may be viewed as the
low energy limit of the interaction shown in Figure~\ref{fig:feyn}
involving the fundamental underlying Yukawa interaction, with strength
$\lambda$, of the scalars to heavy fermions associated to a high
energy scale M. The dimensionless coupling $g_{eff}$ can be understood
as $g_{eff}\equiv \lambda ^2(m_{b}/M)$. Equation~(\ref{eq:lagrangian})
leads, via the exchange of two scalars depicted in Figure~\ref{fig:feyn2}, to the asymptotic potential (i.e. $r\gg m_{\phi}^{-1}$)
\begin {equation}
\label{eq:potyuk}
V\simeq {-g_{eff}^2\over 64\pi^2m_{b}^2}{T \over
r^2}e^{-2m_{\phi}r}
\end {equation}
between two static matter fermions sitting in a scalar heat bath
at temperature $T>T_c$, where $T_c$ is the Bose-Einstein
condensation temperature~\cite{Ferrer:2000hm}.

Below $T_c$, i.e. in the condensed phase, the resulting potential
reads~\cite{Ferrer:2000hm},
\begin {equation}
\label{eq:potcond}
V\simeq  -{g_{eff}^2\over 16\pi m_{b}^2} {q \over
m_{\phi}r}=-{g_{eff}^2\over 48\pi m_{b}^2} {T_c ^2 \over r}
\end {equation}

The effective interaction~(\ref{eq:lagrangian}) is the simplest form describing the exchange of two scalars among baryons. Other forms, like e. g. derivative couplings appropriate for Goldstone bosons, would result in a different $r$ behaviour~\cite{Ferrer:1998ue} of the potentials~(\ref{eq:potyuk})~and~(\ref{eq:potcond}), but the transition to a long range interaction below $T_c$, with the removal of the exponential damping factor, remains unaffected. 

Having mentioned Goldstone bosons, let us, at this point, note that there is an interesting parallel between spontaneous symmetry breaking and Bose-Einstein condensation: both types of systems possess bosonic fields which acquire a constant classical component. This parallel is not exact, however, since the physics of how these constant classical components arise is different (see e.g. ref~\cite{Kapusta:aa}). For instance, the noninteracting charged scalar field cannot exhibit spontaneous symmetry breaking, but it does show the phenomenon of Bose-Einstein condensation and this is indeed the process that occurs in our setting. The scalars that are exchanged in Figure~\ref{fig:feyn2} are {\it not} Goldstone bosons (note e. g. that they have a non-zero mass). Linear couplings are the simplest alternative to describe their interaction with the baryons, and thus our choice in eq.~(\ref{eq:lagrangian}).

\subsection{Influence on the CMB}
Because of this long range dispersion force between baryons, the
Euler equation for baryonic matter in the tightly coupled
photon-baryon fluid gets an extra body force term $k\xi$ (in
Fourier space; k is the wave-number associated to a sub-horizon
sized scale) where the potential $\xi$ obeys the Poisson equation:
\begin {equation}
k^2\xi=-{4\pi \kappa }{(a/a_0)^2} {\rho_b \delta_b}
\end {equation}
with $\kappa \equiv {g_{eff}^2\over 48\pi}{{T_c}^2\over {m_b}^4}$ and
where $a$ is the scale factor, the subscript $0$ refers to
"today", and $\rho_b$ and $\delta_b$ are the baryon energy density
and density contrast, respectively.

Combined with the continuity equation for the photon temperature
fluctuation $\Theta \equiv {\delta T \over T}$ one obtains the
oscillator equation responsible for the acoustic peaks in the
power spectrum of the CMB:
\begin {equation}
\ddot{\Theta}+{\dot{R}\over
1+R}\dot{\Theta}+k^2{\tilde{c}}_s^2\Theta=-{k^2\over
3}\Psi-{\dot{R}\over 1+R}\dot{\Phi}-\ddot{\Phi}
\end {equation}
where $\Psi$ and $\Phi$ are the gravitational potentials
associated to dark matter and $R\equiv {3\rho_{b} \over
4\rho_{\gamma}}$. The derivatives are with respect to conformal
time $\eta$. The key difference with the usual acoustic equation~\cite{Hu:1995em,Hu:2001bc}
is the "speed of sound". Indeed, instead of the proper velocity of
sound $c_{s}^2={1\over 3}{1\over {1+R}}$, we have
${\tilde{c}}_s^2=c_s^2(1-\epsilon_s(k))$ where,
\begin {equation}
\epsilon_s(k)= 3\times 10^4(\Omega_b h^2)^2{12\pi \kappa \over
k^2}({eV/cm^3})
\end {equation}
As in the usual case, the equation can be solved very accurately
with the help of the WKB approximation. The phase of the
oscillations is given by $k\tilde{s}$ with
\begin {equation}
\label{eq:fase}
\tilde{s}=\int \tilde{c}_s d\eta
\end {equation}
i.e. the effective sound horizon.

Modes that verify $k_n=n\pi/\tilde{s}_{rec}$ at recombination are
caught at maxima (compression) or minima (rarefaction) of their
oscillation and render the acoustic peaks in the power spectrum.
Since multipole moments $l\propto k$ approximately,
\begin {equation}
{\delta l_n\over l_n}\simeq {\delta k_n\over k_n}\simeq -\left.\frac{\delta
s}{s}\right|_{rec}
\end {equation}
where $\delta s\equiv \tilde{s}-s$. As a consequence of the
infinite-range interaction, the $l$-position of the peaks will be
shifted (except the first one, see below). Treating
$\epsilon_s(k)$ in equation~(\ref{eq:fase}) as a small perturbation we find:
\begin {equation}
\left.{\delta s\over s}\right\vert_{rec}=-\frac
{1}{2s(\eta_{rec})}\int_{\eta_{ent}}^{\eta_{rec}}\epsilon_s(k)c_s
d\eta
\end {equation}
Note that the above integral starts at $\eta_{ent}$, which is the
time at which the $k_n$-mode enters the horizon. From this moment
onwards, the causal interaction mediated by $2\phi$-exchange is
operative. The integrand depends on time because the speed of
sound changes with time but also because the parameter
$\epsilon_s(k)$ depends on the critical temperature $T_c$ which
decreases as the Universe expands. The integral is easily done
analytically but the explicit expression is cumbersome. If we work
it out a little bit by introducing its numerical value at
recombination, we obtain:
\begin{eqnarray}
\label{eq:deltas}
\fl \left.{\delta s\over s}\right\vert_{rec}&=&-\frac {4.31\times 10^{13}
g_{eff}^2}{(k/0.01Mpc^{-1})^2}\left[1+\left(0.75-7.74R_{ent}\right)
\frac {\sqrt
{0.17+1.17R_{ent}+R_{ent}^2}}{R_{ent}^2}\right.\nonumber\\
\fl             & -&\left. 18.3\ln \frac
{\sqrt{R_{ent}}}{\sqrt{0.17+R_{ent}}+0.41\sqrt{1+R_{ent}}}\right]
\left({0.1eV\over
m_\phi}\right)^2\left({\Omega_bh^2\over 0.02}\right)\left({\Omega_{DM}h^2\over 0.14}\right)
\end{eqnarray}
In equation~(\ref{eq:deltas}) we used the value $R_{rec}=0.46$ and for
$R_{ent}$ we shall use
\begin {equation}
R_{ent}=3.66\left(k/0.01Mpc^{-1}\right)^{-2}\left({\Omega_{DM}h^2\over 0.14}\right)
\end {equation}
which is adequate for scales that enter the horizon after
matter-radiation equality.

Equation~(\ref{eq:deltas}) evaluated at the first peak (i.e. $k\simeq
0.03Mpc^{-1}$) vanishes. It is rightly so, because the first peak
corresponds to a scale that entered the horizon just before
recombination and therefore the dispersion force has had no time
to operate. Higher harmonics, that correspond to smaller scales
and hence had more time to oscillate, give positive shifts
(towards larger $l$'s) because our force effectively reduces the
sound horizon. Finally and obviously, the higher the multipole,
the larger the shift.

\section{A numerical case study}\label{sec:numeric}

The height and location of the acoustic peaks depends on the parameters of the working cosmological model. For instance, the location of the first peak strongly favors a flat Universe with a less than critical baryon density and, combined with the existence of the second peak, one has an estimate of the total cold dark matter density~\cite{Hu:1995em,Hu:2001bc}. Modifications of the cosmological model will show up, thus, in the peak structure, and constraints on such variations can be obtained by means of numerical likelihood analysis. For example, the speed of sound of the dark energy is $c_s \ll 1$ in models with noncanonical kinetic energy density, such as $k$ essence, and densities of the order of $1\%$ can be distinguished from models whose sound speed is near the speed of light~\cite{Erickson:2001bq}. Theories of neutrino mass generation can lead to a uniform shift of the peaks to larger $l$ and an enhanced damping due to a delayed matter-radiation equality~\cite{Chacko:2003dt}. 

As we do not want to endeavor a full analysis of the allowed parameter space in our model, for the sake of definiteness, we take the uncertainty in the second peak position to be $\frac {\delta l_2}{l_2}=1\%$. This figure provides an order of magnitude estimate of present and planned experiments' ability to reconstruct the CMB power spectrum~\cite{Hu:2001bc}. 

Then the third peak moves by a $3\%$. From equation~(\ref{eq:deltas})
we infer that these variations follow for an arbitrarily chosen scalar
mass $m_{\phi}=0.1 eV$and for a coupling $g_{eff}\approx
2.3\times10^{-8}$, which is a very small number. Let us now check the
self-consistency of the picture when we adopt this value for
$g_{eff}$.

The scalar boson gas is made of particles with $+1$ $Q$-charge,
which we denote by $\phi_+$, and anti-particles with $-1$
$Q$-charge denoted by $\phi_-$. To easy notation, however, we
shall display the charge sign only when strictly necessary. Let the net
total charge be positive (this is purely conventional) and
therefore there is an excess of particles $\phi_+$  over
antiparticles $\phi_-$ in the Universe. At the very early epochs
of the Universe, say at $T\gg M$, when the degrees of freedom
associated to the high energy scale M (e.g. fermions $\chi$ with charge Q
in Figure~\ref{fig:feyn}) are fully relativistic, our scalars are in thermal
contact with all relativistic species, the density is very high,
and the gas of scalars is condensed with most of the charge
sitting in the condensate. The condensed phase consists
overwhelmingly of particles $\phi_+$ of zero momentum. They are
the bulk of dark matter. Since charge $Q$ in any co-moving volume
is conserved, the charge density $q$ is bound to be:
\begin {equation}
q=1.5\times 10^4(1+z)^3\left(\frac
{\Omega_{DM}h^2}{0.14}\right)\left(\frac
{0.1eV}{m_\phi}\right)cm^{-3}
\end {equation}
Although charge in excited states is a smallish fraction $O(m_\phi
/T)$ of the charge in the condensed phase, the number density of
scalars and anti-scalars occupying excited states is comparable to the
other relativistic species at those large temperatures. Later on, when
temperature has dropped below the energy scale M, and consequently
Q-fermions are extinct, our scalars might still interact with baryons
through elastic $\phi N$-scattering ($N$ stands for nucleon) where the
$\phi$-interaction rate $\Gamma \sim n_{b}\langle \sigma v \rangle$
contains the properly averaged cross section $\times$ relative
velocity:

\begin {equation}
\fl \langle \sigma  v \rangle= \frac {1}{2\langle p_{\phi}^0\rangle}
\frac {1}{2\langle p_{b}^0\rangle}\int \frac
{d^3p_{\phi}'}{(2\pi)^32p_{\phi}'^0}\frac
{d^3p_{b}'}{(2\pi)^32p_{b}'^0}|\textit{M}|^2(1+f_{BE})(2\pi)^4\delta^4(P-P')
\end {equation}
In this expression $|\textit{M}|^2$ is the spin averaged
transition matrix element squared and we have included the
spontaneous plus stimulated emission factor $(1+f_{BE})$, where
$f_{BE}=[\exp \left(p_{\phi}'^0-\mu\right)/T-1]^{-1}$ is the Bose-Einstein
distribution function ($\mu$ being the chemical potential). This
factor weighs the population of final particle phase space in a
boson medium. The calculation of the interaction rate for thermal
scalars (i.e. those with nonzero momentum) gives a value, at
$T\sim 1GeV$, comparable to the Hubble expansion rate but drops
precipitously beyond $N\bar{N}$ annihilation due to the massive
loss of scatterers. The calculation of the interaction rate for
the zero momentum scalars in the condensate is not as
straightforward.

  Indeed, the final scalar phase space
integral in the previous equation does not properly account for
the sum over states in the condensed regime. The zero modes are
not included in the sum. This is not a problem in the
non-degenerate case because charge is thinly distributed over the
states and the ground state gets only an infinitesimal share of
the total charge. However, in a condensed Bose-Einstein gas a
macroscopic fraction of the charge sits in the ground state. In
our case, the ground state is populated essentially by particles
$\phi_+$. Actually, the integral (evaluated at fixed chemical
potential $\mu=m_\phi$) gives then the contribution of the excited
states to the phase space. We shall estimate the stimulated
emission contribution to the interaction rate to be:

\begin {equation}
\label{eq:gammastim}
\Gamma_{stim}\approx n_b\langle \sigma^* v \rangle \frac {n}{n^*}
\end {equation}
where $\langle \sigma^* v \rangle$ is explicitly given by\footnote
{the star in $\sigma ^*$ is a reminder that we are summing only
over excited final states}:
\begin {equation}
\label{eq:sigmastar}
\fl \langle \sigma^* v \rangle=\frac {1}{2 m_\phi} \frac {1}{2
m_{b}}\int \frac {d^3p_{\phi}'}{(2\pi)^32p_{\phi}'^0}\frac
{d^3p_{b}'}{(2\pi)^32p_{b}'^0}|\textit{M}|^2\frac
{1}{\exp(p_\phi'^0-m_\phi)/T-1} (2\pi)^4\delta^4(P-P')
\end {equation}
where we have omitted the spontaneous emission term which in this
case is negligible compared to the stimulated emission term, and
the fraction $n/n^*$ of the total number of particles over the
particles in excited states can be approximated by:
\begin {equation}
\frac {n}{n^*}=\frac {n_{k=0}+n^*}{n^*}\approx 1+\frac
{q_{k=0}}{n^*}=1+\frac {q\left [1-\left(\frac
{T}{T_c}\right)^2\right]}{\zeta (3)T^3/\pi ^2}
\end {equation}
 In equation~(\ref{eq:sigmastar}) we took into account that the rates are to be evaluated
in a co-moving frame at rest with respect to the Hubble flux. In
this frame the scalars in the condensate are motionless and the
baryons are non-relativistic in the relevant temperature span
($T\sim 1GeV$ and below). The energy-momentum deltas can be used
to crack down the phase space integrals and to obtain a much
simpler expression that reads:
\begin {equation}
\label{eq:sigmacrack}
\fl \langle \sigma^* v \rangle=\frac {1}{2 m_\phi} \frac {1}{2
m_{b}}\int \frac {dz}{4\pi}\delta (z-1)|\textit{M}|^2\frac
{1}{\exp(p_\phi'^0-m_\phi)/T-1}\frac
{p'_b}{p_\phi'^0+p_b'^0\left(1+\frac {p_b}{p_b'}z\right)}
\end {equation}
with $p_b=|\textbf{p}_b|$, $p_b'=|\textbf{p}_b'|$ and z is the
cosine of the "laboratory" scattering angle. The kinematics of the
scattering process corresponds to an extremely glancing collision
as required by energy-momentum conservation and the delta function
in equation~(\ref{eq:sigmacrack}) describes approximately the fact that $1\geq z^2
\geq 1-\left(m_\phi/m_b \right)^2$. In equation~(\ref{eq:sigmacrack}), $p_b'$ and
$p_\phi'^0$ are fixed by energy-momentum conservation:
\begin {equation}
p_b'\simeq p_b\left(1-\frac {m_\phi}{m_b}\right)
\end {equation}
and

\begin {equation}
 p_\phi'^0\simeq m_\phi
\left(1+\frac {1}{2}\frac {p_b^2}{m_b^2}\right)
\end {equation}
Collecting the different pieces and introducing them in equation~(\ref{eq:sigmacrack}), we obtain:
\begin {equation}
\label{eq:sigmastarv}
\langle \sigma^* v \rangle=\frac
{\sqrt{3}}{16\pi}g_{eff}^2m_\phi^{-2}\left(\frac
{T}{m_b}\right)^{1/2}
\end {equation}
where we used $p_b\simeq \sqrt{3m_bT}$ as dictated by the
equipartition theorem.

We get the final form for the interaction rate of scalars in the
condensate by plugging the equation above into equation~(\ref{eq:gammastim}) where
the factor $n/n^*$ is about $70\left(\frac
{\Omega_{DM}}{0.14}\right)\left(\frac {0.1 eV}{m_\phi}\right)\left(\frac {T}{T_{\phi(k \neq 0)}}\right)^3$ in
this temperature regime ($T\sim 1GeV$ and below). Even below
anti-baryon extinction, when the baryon number density $n_b$
 is $1.22\times 10^{-10}T^3$, the resulting rate is many orders of magnitude larger than the Hubble
 rate:
\begin {equation}
\fl \frac {\Gamma}{H}\approx 4\times 10^{13}\left(\frac
{g_{eff}}{2.3\times10^{-8}}\right)^2\left(\frac
{0.1eV}{m_\phi}\right)^3\left(\frac
{1GeV}{m_b}\right)^{1/2}\left(\frac
{\Omega_{DM}h^2}{0.14}\right)^{1/2}\left(\frac
{T}{1GeV}\right)^{3/2}
\end {equation}
Hence, while thermal scalars are already decoupled at temperatures
below or about $1GeV$, the zero modes in the scalar condensate are
in full thermal contact with the cosmic plasma. Though decoupled,
thermal scalars keep their equilibrium distributions with momenta
redshifting as the Universe expands. And hence they keep a common
temperature with the condensate bosons that are tightly coupled to
baryons and photons. A common temperature is maintained only until anti-baryon extinction takes place. By
then photons are reheated with respect to thermal scalars and the
condensate, still coupled to the baryon-photon fluid, shares its
temperature. Actually the zero modes in the condensate keep in
thermal contact with the baryons well beyond matter-radiation
equality (see below). A further widening of the temperature gap
between thermal scalars on the one hand and photons and the
condensate on the other hand occurs at two more stages: after $\mu^+\mu^-$ and later after $e^+e^-$ annihilation.
Hence at nucleosynthesis the ratio of temperatures is:
\begin {equation}
\label{eq:reldof}
\frac {T_ {\phi (k\neq 0)}}{T}=\left(\frac
{rel.d.o.f.\; T \leq m_e}{rel.d.o.f.\; T \geq 1GeV}\right)^{1/3}\sim 0.3 
\end {equation}
Since the effective number of degrees of freedom associated to the
thermal scalars at nucleosynthesis is proportional to this
fraction to the fourth power, there is clearly no conflict with
primordial helium yields~\cite{Fields:2004cb}. We turn next to the question as to when
does the condensate loose thermal contact with the baryon-photon
plasma.

In a matter dominated Universe the expansion rate is given by:
\begin {equation}
H=\sqrt{\frac{8\pi G}{3}}\rho_M^{1/2}
\end {equation}
$\rho_M$ being the non-relativistic matter density.

  Decoupling of the particles in the boson condensate from the
baryon-photon fluid happens when their interaction rate roughly
equals the cosmic expansion rate.  Actual numerical comparison of
both rates renders:
\begin {equation}
\fl \frac {\Gamma}{H}\sim \left(\frac
{g_{eff}}{2.3\times10^{-8}}\right)^2\left(\frac
{0.1eV}{m_\phi}\right)^3\left(\frac
{1GeV}{m_b}\right)^{1/2}\left(\frac
{\Omega_{DM}h^2}{0.14}\right)\left(\frac
{0.15}{\Omega_Mh^2}\right)^{1/2}\left(\frac {T}{0.016eV}\right)^2
\end {equation}
This is a rough estimate that
determines the moment of condensate decoupling to happen about a factor $10-20$ below the recombination temperature. Here we took into account that scalars in the condensate and thermal scalars do not share the same temperature at these later stages. Indeed, in this temperature regime we replaced $T_{\phi(k \neq 0)}$ for $T$ in equation~(\ref{eq:sigmastarv}), we used $n/n^* \sim 70\left(\frac
{\Omega_{DM}}{0.14}\right)\left(\frac {0.1 eV}{m_\phi}\right)\left(\frac {T}{T_{\phi(k \neq 0)}}\right)^3$, and we related both $T$ and $T_{\phi(k \neq 0)}$ via equation~(\ref{eq:reldof}). 

The charge density at decoupling is still about a factor of five above
critical. So the transition to an uncondensed phase is yet to
happen. However, since scalars are non-interacting and thermally
isolated from ordinary matter ever since decoupling took place,
scalars will maintain the equilibrium distributions they had at
decoupling. The zero modes, i.e. the condensate, will therefore keep
in the lowest energy state beyond the critical point because no
interaction can pump them out of the ground state. As a consequence,
baryons in the cosmic scalar background today, i.e. at zero redshift,
should feel the long-range two body exchange forces under
scrutiny. They are extremely feeble. In fact, being proportional to
the present cosmic Q-charge density and to $g_{eff}^2\sim 5\times
10^{-16}$ (see Equation~(\ref{eq:potcond})), they turn out to be
almost a million times weaker than the force of gravity. Hence for
this particular numerical example, they would go unnoticed by local
gravity experiments~\cite{Damour:tw}, although, in general, couplings
between the dark and the visible sectors can have observable
consequences on structure
formation~\cite{Tocchini-Valentini:2001ty}.

\section{Conclusion} \label{sec:conclusio}
We have entertained the possibility that the dark matter is composed by a Bose-Einstein condensate of light scalar particles coupled to the baryonic component of the Universe. The long-range forces among baryons caused by the pair exchange of dark matter particles in the presence of the condensate gives rise to a shift in the position of the acoustic peaks in CMB power spectrum above the first one. 

For a particular choice of the parameters in our
model, a consistent cosmic evolution results that would show up as
variations in the peak locations of the order of a few percent. Since
our primary aim was to display the consequences of Bose-Einstein condensation
of a relativistic scalar gas in a physical setting, we do not find it justified to exhaustively explore the full $m_{\phi}-g_{eff}$ parameter space. Other choices of the parameters could lead to a completely different, and even inconsistent, cosmic evolution. For example, the transition to a non-condensed phase could take place before the thermal decoupling, and we would be left with light scalar dark matter particles thinly populating the states above the ground state with no trace of the primordial condensate before structure formation occurs. Also, tests of the
gravitational inverse-square law~\cite{Hoyle:2004cw} or Casimir force
measurements~\cite{Decca:2003td}, could be sensitive to parts of the
$m_{\phi}-g_{eff}$ parameter space different from the ones studied. 

In the particular case considered, we find that the particles in the condensate will remain in the lowest energy state beyond the critical point. Once decoupled, the particles in the condensate are in practice isolated from the ones in the thermally excited states and momentum redshifting can not cause any flow between this two populations.

\ack
Work partially supported by the CICYT Research Project FPA2002-0064, the EU network on Supersymmetry and the Early Universe HPRN-CT-2000-00152 and the DURSI Research Project 2001SGR00188. FF thanks the IFAE for hospitality, his research is supported by the Leverhulme Trust.



\section*{References}

\begin{thebibliography}{99}

\bibitem{Feinberg:ps}
G.~Feinberg, J.~Sucher and C.~K.~Au,
Phys.\ Rept.\  {\bf 180}, 83 (1989).

\bibitem{Grifols:zz}
J.~A.~Grifols and S.~Tortosa,
Phys.\ Lett.\ B {\bf 328}, 98 (1994)
[arXiv:hep-ph/9404249].
J.~A.~Grifols, E.~Masso and R.~Toldra,
Phys.\ Lett.\ B {\bf 389}, 563 (1996)
[arXiv:hep-ph/9606377].

\bibitem{Ferrer:1998ue}
F.~Ferrer and J.~A.~Grifols,
Phys.\ Rev.\ D {\bf 58}, 096006 (1998)
[arXiv:hep-ph/9805477].

\bibitem{Ferrer:1998ju}
F.~Ferrer, J.~A.~Grifols and M.~Nowakowski,
Phys.\ Lett.\ B {\bf 446}, 111 (1999)
[arXiv:hep-ph/9806438].
F.~Ferrer and J.~A.~Grifols,
Phys.\ Lett.\ B {\bf 460}, 371 (1999)
[Erratum-ibid.\ B {\bf 511}, 319 (2001)]
[arXiv:hep-ph/9904394].
F.~Ferrer, J.~A.~Grifols and M.~Nowakowski,
Phys.\ Rev.\ D {\bf 61}, 057304 (2000)
[arXiv:hep-ph/9906463].

\bibitem{casimir}
H.~B.~G.~Casimir and P.~Polder,
Phys.\ Rev.\ {\bf 73}, 360 (1948).

\bibitem{sucher}
G.~Feinberg and J.~Sucher,
Phys. Rev. {\bf 166}, 1638 (1968).
S.~D.~H.~Hsu and P.~Sikivie,
Phys.\ Rev.\ D {\bf 49}, 4951 (1994)
[arXiv:hep-ph/9211301].
E.~Fischbach,
Annals Phys.\  {\bf 247}, 213 (1996)
[arXiv:hep-ph/9603396].

\bibitem{Barton:2001ej}
G.~Barton,
Phys.\ Rev.\ A {\bf 64}, 032102 (2001).
C.~J.~Horowitz and J.~Pantaleone,
Phys.\ Lett.\ B {\bf 319}, 186 (1993)
[arXiv:hep-ph/9306222].

\bibitem{Ferrer:1998rw}
F.~Ferrer and M.~Nowakowski,
Phys.\ Rev.\ D {\bf 59}, 075009 (1999)
[arXiv:hep-ph/9810550].

\bibitem{Ellis:1983ew}
J.~R.~Ellis, J.~S.~Hagelin, D.~V.~Nanopoulos, K.~A.~Olive and M.~Srednicki,
Nucl.\ Phys.\ B {\bf 238}, 453 (1984).
H.~Goldberg,
Phys.\ Rev.\ Lett.\  {\bf 50}, 1419 (1983).

\bibitem{Matos:2000ki}
T.~Matos, F.~S.~Guzman and D.~Nunez,
Phys.\ Rev.\ D {\bf 62}, 061301 (2000)
[arXiv:astro-ph/0003398].
T.~Matos and L.~A.~Urena-Lopez,
arXiv:astro-ph/0406194.


\bibitem{Jean:2003ci}
P.~Jean {\it et al.},
Astron.\ Astrophys.\  {\bf 407}, L55 (2003)
[arXiv:astro-ph/0309484].

\bibitem{Boehm:2003bt}
C.~Boehm, D.~Hooper, J.~Silk and M.~Casse,
Phys.\ Rev.\ Lett.\  {\bf 92}, 101301 (2004)
[arXiv:astro-ph/0309686].
D.~Hooper, F.~Ferrer, C.~Boehm, J.~Silk, J.~Paul, N.~W.~Evans and M.~Casse,
arXiv:astro-ph/0311150.





\bibitem{Ferrer:2000hm}
F.~Ferrer and J.~A.~Grifols,
Phys.\ Rev.\ D {\bf 63}, 025020 (2001)
[arXiv:hep-ph/0001185].

\bibitem{Consoli:2003jv}
M.~Consoli and E.~Costanzo,
Eur.\ Phys.\ J.\ C {\bf 33}, 297 (2004)
[arXiv:hep-ph/0311317].

\bibitem{Sannino:2003mt}
F.~Sannino and K.~Tuominen,
Phys.\ Rev.\ D {\bf 68}, 016007 (2003)
[arXiv:hep-ph/0303167].
F.~Sannino and K.~Tuominen,
arXiv:hep-ph/0305004.



\bibitem{Bertone:2004pz}
G.~Bertone, D.~Hooper and J.~Silk,
arXiv:hep-ph/0404175.
C.~Munoz,
arXiv:hep-ph/0309346.

\bibitem{Krauss:2004iq}
L.~M.~Krauss,
arXiv:astro-ph/0406673.

\bibitem{Blanchard:2003du}
A.~Blanchard, M.~Douspis, M.~Rowan-Robinson and S.~Sarkar,
Astron.\ Astrophys.\  {\bf 412}, 35 (2003)
[arXiv:astro-ph/0304237].

\bibitem{Peebles:1987ek}
N.~Weiss,
Phys.\ Lett.\ B {\bf 197}, 42 (1987).
C.~Wetterich,
Nucl.\ Phys.\ B {\bf 302}, 668 (1988).
P.~J.~E.~Peebles and B.~Ratra,
Astrophys.\ J.\  {\bf 325}, L17 (1988).
P.~G.~Ferreira and M.~Joyce,
Phys.\ Rev.\ D {\bf 58}, 023503 (1998)
[arXiv:astro-ph/9711102].
R.~R.~Caldwell, R.~Dave and P.~J.~Steinhardt,
Phys.\ Rev.\ Lett.\  {\bf 80}, 1582 (1998)
[arXiv:astro-ph/9708069].
G.~Huey, L.~M.~Wang, R.~Dave, R.~R.~Caldwell and P.~J.~Steinhardt,
Phys.\ Rev.\ D {\bf 59}, 063005 (1999)
[arXiv:astro-ph/9804285].






\bibitem{Enqvist:2003gh}
K.~Enqvist and A.~Mazumdar,
Phys.\ Rept.\  {\bf 380}, 99 (2003)
[arXiv:hep-ph/0209244].
M.~Dine and A.~Kusenko,
Rev.\ Mod.\ Phys.\  {\bf 76}, 1 (2004)
[arXiv:hep-ph/0303065].
R.~Allahverdi, R.~Brandenberger and A.~Mazumdar,
arXiv:hep-ph/0407230.

\bibitem{Bassett:2002fe}
B.~A.~Bassett, M.~Kunz, D.~Parkinson and C.~Ungarelli,
Phys.\ Rev.\ D {\bf 68}, 043504 (2003)
[arXiv:astro-ph/0211303].

\bibitem{Gubser:2004uh}
G.~R.~Farrar and P.~J.~E.~Peebles,
Astrophys.\ J.\  {\bf 604}, 1 (2004)
[arXiv:astro-ph/0307316].
S.~S.~Gubser and P.~J.~E.~Peebles,
arXiv:hep-th/0402225.
S.~S.~Gubser and P.~J.~E.~Peebles,
arXiv:hep-th/0407097.



\bibitem{Haber:fg}
H.~E.~Haber and H.~A.~Weldon,
Phys.\ Rev.\ Lett.\  {\bf 46}, 1497 (1981);
Phys.\ Rev.\ D {\bf 25}, 502 (1982).

\bibitem{Kapusta:aa}
J.~I.~Kapusta,
Phys.\ Rev.\ D {\bf 24} (1981) 426.

\bibitem{Bernstein:kf}
J.~Bernstein and S.~Dodelson,
Phys.\ Rev.\ Lett.\  {\bf 66}, 683 (1991).

\bibitem{Hu:1995em}
See e.g. W.~Hu and N.~Sugiyama,
Phys.\ Rev.\ D {\bf 51}, 2599 (1995)
[arXiv:astro-ph/9411008].
W.~Hu, N.~Sugiyama and J.~Silk,
Nature {\bf 386}, 37 (1997)
[arXiv:astro-ph/9604166].
V.~Mukhanov,
arXiv:astro-ph/0303072.


\bibitem{Hu:2001bc}
W.~Hu and S.~Dodelson,
Ann.\ Rev.\ Astron.\ Astrophys.\  {\bf 40}, 171 (2002)
[arXiv:astro-ph/0110414].

\bibitem{Erickson:2001bq}
J.~K.~Erickson, R.~R.~Caldwell, P.~J.~Steinhardt, C.~Armendariz-Picon and V.~Mukhanov,
Phys.\ Rev.\ Lett.\  {\bf 88}, 121301 (2002)
[arXiv:astro-ph/0112438].

\bibitem{Chacko:2003dt}
Z.~Chacko, L.~J.~Hall, T.~Okui and S.~J.~Oliver,
arXiv:hep-ph/0312267.
S.~Bashinsky and U.~Seljak,
Phys.\ Rev.\ D {\bf 69}, 083002 (2004)
[arXiv:astro-ph/0310198].


\bibitem{Fields:2004cb}
See e.g. B.~Fields and S.~Sarkar,
arXiv:astro-ph/0406663.


\bibitem{Damour:tw}
T.~Damour, G.~W.~Gibbons and C.~Gundlach,
Phys.\ Rev.\ Lett.\  {\bf 64} (1990) 123.


\bibitem{Tocchini-Valentini:2001ty}
L.~Amendola,
Phys.\ Rev.\ D {\bf 62}, 043511 (2000)
[arXiv:astro-ph/9908023].
D.~Tocchini-Valentini and L.~Amendola,
Phys.\ Rev.\ D {\bf 65}, 063508 (2002)
[arXiv:astro-ph/0108143].
C.~Boehm, P.~Fayet and R.~Schaeffer,
Phys.\ Lett.\ B {\bf 518}, 8 (2001)
[arXiv:astro-ph/0012504].
C.~Boehm, A.~Riazuelo, S.~H.~Hansen and R.~Schaeffer,
Phys.\ Rev.\ D {\bf 66}, 083505 (2002)
[arXiv:astro-ph/0112522].




\bibitem{Hoyle:2004cw}
C.~D.~Hoyle, D.~J.~Kapner, B.~R.~Heckel, E.~G.~Adelberger, J.~H.~Gundlach, U.~Schmidt and H.~E.~Swanson,
arXiv:hep-ph/0405262.
E.~G.~Adelberger, E.~Fischbach, D.~E.~Krause and R.~D.~Newman,
Phys.\ Rev.\ D {\bf 68}, 062002 (2003).
E.~G.~Adelberger, B.~R.~Heckel and A.~E.~Nelson,
Ann.\ Rev.\ Nucl.\ Part.\ Sci.\  {\bf 53}, 77 (2003)
[arXiv:hep-ph/0307284].
E.~Fischbach and D.~E.~Krause,
Phys.\ Rev.\ Lett.\  {\bf 83}, 3593 (1999)
[arXiv:hep-ph/9906240].

\bibitem{Decca:2003td}
R.~S.~Decca, E.~Fischbach, G.~L.~Klimchitskaya, D.~E.~Krause, D.~L.~Lopez and V.~M.~Mostepanenko,
Phys.\ Rev.\ D {\bf 68}, 116003 (2003)
[arXiv:hep-ph/0310157].











\end{thebibliography}
\end{document}